# Demand Response Requirements from the Cultural, Social, and Behavioral Perspectives

Mohammadreza Shekari[1], Hamidreza Arasteh[2], Alireza Sheikhi Fini[3], Vahid Vahidinasab[4]

[1] Power systems group, Department of Electrical and Computer Engineering, Tarbiat Modares University, Tehran, Iran, m_shekari@modares.ac.ir
[2] Power Systems Operation and Planning Research Department, Niroo Research Institute, Tehran, Iran, harasteh@nri.ac.ir
[3] Power Systems Operation and Planning Research Department, Niroo Research Institute, Tehran, Iran, asheikhi@nri.ac.ir
[4] Department of Engineering, School of Science and Technology, Nottingham Trent University, Nottingham NG11 8NS, UK, vahid.vahidinasab@ntu.ac.uk

**Abstract:** Demand-side response programs which also called Demand Response (DR) are interesting ways to attract consumers' participation in order to improve electric consumption patterns. DR programs motivate customers to change consumption patterns in response to price changes. This could be done by paying incentives or considering penalties either when wholesale market prices are high or when network reliability is at risk. The overall purpose of implementing DR programs is to improve reliability of the networks, while reducing the operation costs. Successful implementation of these DR programs requires prerequisites, which without them, there is no guarantee of success. Different disciplines have provided various scientific solutions for creating optimal power consumption behavior in customers (such as utilizing intelligent home appliances and big data techniques). Although each of these solutions might be efficient and effective, they could not cover all the aspects of the solutions. The results of studies conducted by many researchers show that in addition to the technical and economic issues, social, cultural, and behavioral factors are of great importance in DR implementation. Therefore, in this paper, these aspects, as one of the vital requirements for better implementation of DR, are investigated and analyzed in detail.

## 1. Introduction

Before the 1973 oil crisis, the issue of energy was not considered as a crucial concern [1]. As crude oil and energy prices rose, the trend in energy consumption changed dramatically, and oil-importing countries moved toward more energy savings and better use of available energy [2]. Besides the technical aspects, the role of the consumers has gained high attention. Therefore, various researches were started in order to identify an effective solution based on the consumers' behavior to optimize the energy consumption patterns. Demand response (DR) programs are highly interested worldwide duo to their potential to engage end-users in the electricity delivery chain [3], [4]. DR is defined as the ability of the customers to change their consumption patterns in response to the signals from the market/system [5], [6]. Since DR implementation is related to the customers' decision to be responsive or not, it is a multi-disciplinary problem, including various perspectives such as economic, social, and cultural aspects besides the electrical points of view. In this regard, the study of humans, their specific values, beliefs and behaviors based on the economic, social and cultural assets became important. Furthermore, how communities respond to different problems and how environmental beliefs and values are related to the ways that they are using the environment, have been considered as a significant research area [7].

There are different scientific solutions in response to the question of how to change the customers' energy consumption pattern in an optimal way. For example, technical experts believe that technological improvement can optimize energy consumption [8]. Economists, mainly emphasize that energy consumption can be controlled by increasing the prices of energy carriers. Although each of these solutions is efficient and effective, they are not sufficient. The results of studies conducted by various researchers in different parts of the world show that in addition to the technical and economic issues, other factors such as social, cultural, and behavioral ones are also considerable.

In order to study the energy consumption behaviors, it is important to identify the effective factors on the load pattern. The role of cultural factors (such as attitude, education level and the awareness of the consumers about the issues), and social factors (such as family, friends, and social networks) are considerable. By determining the influential factors, the necessary training can be provided to the consumers in order to optimize their consumption pattern by means of cultural transformation. In this regard, it is necessary to identify the social and cultural parameters affecting the behavior of the customers, as well as their roles and impacts [9].

This paper identifies the parameters that could affect electricity consumption i.e., human's behavior. In other words, this paper will examine the factors affecting the electricity consumption behavior and requirements for implementing the DR programs. The main results of the literature review in the case of behavioral aspects are presented in Table 1. The following information was obtained from a study of 54 references over the last 20 years. As the table shows, most studies have been conducted on residential customers. The most important factors that have been studied are the social and behavioral factors. Only four references have considered all the factors. Also, it could be observed that environmental and cultural factors need more attention in future research.

The rest of the paper is organized as follows. In section 2, a theoretical framework is introduced to study the behavioral aspects. Section 3 dedicated to the research hypotheses and some important concepts, and afterward, an analytical model is presented to show the relationship between the proposed behavioral factors. The cultural, social, and behavioral requirements of the DR are presented in Section 4. The relevant solutions and suggestions are also provided in this section. Finally, the conclusions are provided in section 5.

## 2. Theoretical Framework and Concept

In this section, the main concept and sociological perspective of the paper are presented. Moreover, summarized topics and final requirements for implementing DR programs are suggested. Examining the relevant theories (as a general guide for answering the questions and achieving the mentioned goals) shows the use of the three categories of theories as follows:

- Introducing customers to electricity consumption management;
- Theories related to consumption and lifestyle;
- The new environmental paradigm and its value and knowledge.

*2.1. Introducing customers to electricity consumption management*

Economic development and industrial prosperity are not possible without the development of the electricity industry. Lack of electricity and forced blackouts have caused irreparable damage to the world's economy, especially in recent years. If this trend continues, the recession will be inevitable [10].

**Table 1.** Taxonomy of the studies from 5 different aspects

| Ref. | Cultural | Social | Behavioral | Economic | Environmental | Type of customers |
|------|----------|--------|------------|----------|---------------|-------------------|
| [9]  | ✓ |   | ✓ |   | ✓ | Urban citizens |
| [11] | ✓ |   |   | ✓ |   | Residential |
| [12] |   | ✓ |   | ✓ |   | Residential |
| [13] |   | ✓ |   | ✓ | ✓ | Residential |
| [14] |   | ✓ | ✓ |   |   | Industrial |
| [15] |   | ✓ | ✓ |   |   | Residential (Schools) |
| [16] |   | ✓ |   | ✓ |   | Residential |
| [17] | ✓ | ✓ |   |   |   | Rural citizens |
| [18] |   | ✓ |   | ✓ |   | Industrial |
| [19] | ✓ |   |   | ✓ |   | Industrial & Residential |
| [20] | ✓ |   |   | ✓ |   | Urban citizens |
| [21] |   | ✓ |   |   | ✓ | Urban citizens |
| [22] | ✓ | ✓ |   |   | ✓ | Industrial |
| [23] | ✓ | ✓ | ✓ | ✓ |   | Residential & Urban citizens |
| [24] |   | ✓ | ✓ |   |   | Residential |
| [25] | ✓ | ✓ | ✓ |   | ✓ | Industrial |
| [26] |   |   |   | ✓ |   | Residential |
| [27] |   | ✓ | ✓ |   | ✓ | Urban citizens |
| [28] |   |   | ✓ |   | ✓ | Residential |
| [29] | ✓ | ✓ | ✓ |   |   | Residential (Schools) |

| Ref. | Cultural | Social | Behavioral | Economic | Environmental | Type of customers |
|---|---|---|---|---|---|---|
| [30] | ✓ |  |  | ✓ | ✓ | Urban citizens |
| [31] |  |  | ✓ | ✓ |  | Residential |
| [32] | ✓ | ✓ |  | ✓ |  | Residential |
| [33] |  | ✓ | ✓ |  |  | Industrial |
| [34] |  | ✓ | ✓ | ✓ |  | Residential |
| [35] | ✓ | ✓ | ✓ | ✓ | ✓ | Residential |
| [36] |  | ✓ | ✓ |  |  | Residential |
| [37] |  | ✓ | ✓ |  |  | Residential |
| [38] |  | ✓ |  | ✓ | ✓ | Residential |
| [39] | ✓ |  |  | ✓ | ✓ | Country citizens |
| [40] |  | ✓ | ✓ |  |  | Residential |
| [41] |  |  |  | ✓ |  | Residential |
| [42] | ✓ |  | ✓ |  |  | Rural citizens |
| [43] | ✓ |  |  |  |  | Residential |
| [44] |  | ✓ | ✓ | ✓ |  | Residential & Urban citizens |
| [45] |  |  | ✓ |  |  | Industrial & Residential |
| [46] |  | ✓ |  |  | ✓ | Residential |
| [47] |  | ✓ | ✓ |  |  | Industrial & Residential |
| [48] |  |  | ✓ |  | ✓ | Industrial |
| [49] | ✓ | ✓ | ✓ |  |  | Residential |
| [50] | ✓ |  | ✓ | ✓ |  | Industrial & Residential |
| [51] |  | ✓ | ✓ | ✓ |  | Industrial & Residential |
| [52] |  | ✓ | ✓ | ✓ | ✓ | Country citizens |
| [53] | ✓ | ✓ | ✓ | ✓ | ✓ | Industrial & Residential |
| [54] |  |  | ✓ | ✓ | ✓ | Industrial & Country citizens |
| [55] |  |  | ✓ | ✓ |  | Industrial & Residential |

**Table 1.** Taxonomy of the studies from 5 different aspects (*continued*)

| Ref. | Cultural | Social | Behavioral | Economic | Environmental | Type of customers |
|---|---|---|---|---|---|---|
| [56] |  | ✓ | ✓ | ✓ | ✓ | Industrial |
| [57] | ✓ | ✓ | ✓ | ✓ | ✓ | Residential |
| [58] |  | ✓ |  | ✓ |  | Industrial & Country citizens |
| [59] |  | ✓ |  | ✓ | ✓ | Industrial |
| [60] | ✓ | ✓ | ✓ |  |  | Residential |
| [61] | ✓ | ✓ |  |  |  | Residential |
| [62] | ✓ | ✓ | ✓ | ✓ | ✓ | Industrial & Residential |
| Percentage | 40.7 | 64.8 | 61.1 | 53.7 | 37.0 | - |

Consumption management is one of the important tools utilized by power utility companies or large consumers to reduce consumption peak and smooth the load curve, which increases the efficiency of electricity consumption and reduces the cost. The operating companies are gained from consumption management methods due to eliminating power shortages during peak hours and improving the load factor provided by them. Also, large consumers, such as factories, benefit from the numerous economic, technical, and services enabled by consumption management tools. One way to improve the current situation is to implement Demand Side Management policies, particularly DR programs [10].

Utility companies can also teach appropriate consumption patterns to their big customers and encourage them to use these methods. Utilities utilize different consumption management methods. For example, they offer more expensive electricity in high-consumption months (such as summer) and cheaper electricity in low-consumption months. Large companies and factories could plan their annual leave and vacation during these months in an appropriate way [10]. Owners of large industrial companies can also easily reduce the cost of electricity consumption by purchasing demand-side management programs. These programs are easy to access and use. Universities and computer companies are developing such programs.

*2.2. Theories related to consumption and lifestyle*

Like other sciences, every term of social sciences is understandable in its conceptual context. Thus, for example, the term "lifestyle", as one of the terms of social sciences, is directly related to a set of concepts such as culture and society, objective and subjective culture, form and content, behavior (attitude, value, and norm), and social classification. Without understanding these relationships, a proper understanding of lifestyle and related theories will not be achieved. Lifestyle applications often have generalized concepts and are used for various cases. Sometimes it is mistakenly referred to as culture and class. So, we have to look for the exact meaning that shows the relationship of lifestyle with other phenomena and concepts [63].

There are two interpretations of the concept of "lifestyle" in sociological literature. One relates to the 1920s, in which lifestyle represented individuals' wealth and social status and was often used as an indicator of social class. The second is a new social form that only makes sense in modernity changes and the growth of the consumerist culture. Therefore, from this point of view, lifestyle is a way to define people's values, attitudes, and behaviors and its importance for social analysis are increasing more and more [64].

Usually, lifestyle has been defined by reference to consumption as one of the primary phenomena. Considering the importance of the consumption concept in this paper and its relationship with lifestyle, the opinions of prominent experts who have studied the idea of consumption and lifestyle from a sociological perspective have been examined in this section.

*2.3. The new environmental paradigm: value and knowledge*

Modifying the energy consumption pattern and its procedures has a significant impact on the amount of fuel used to produce energy, which leads to a proper decarbonized environment. Therefore, by examining theories related to the environment and modifying the pattern of energy consumption, it is possible to identify cultural and social factors affecting DR programs. At the beginning of the 21st century, environmental issues have received much more attention than ever before because human societies' damage to the environment is more significant than ever before [65]. Some environmental researchers believe that most of the environmental problems came from human actions. Human abuse of the environment has been identified as the most important threat to the world in the long run. Improper human behavior causes harmful and irreversible changes to the environment. If these problems not be tackled, there will no guarantee to have a sustainable world in the future [66].

Environmental sociologists are very interested in understanding and predicting the environmental behaviors of individuals in a society. One of the criteria for predicting environmental behaviors is attitude. Attitude theory implies that behaviors always have intention and purpose. This theory focuses on how people make choices and decisions in specific situations [65]. The environmental approach addresses issues such as the conservation of wildlife and natural resources, plants, forests, and native animals. This approach also efforts to address environmental problems at the macro level.

There are various sociological theories about why and how people will change their behaviors towards the environment. The new environmental paradigm is one of the important ones [65]. The separation logic (human separation from nature) claimed that the relationship between humankind and the environment is sociologically insignificant. From this perspective, human beings are separate from nature due to the superior power and having a culture. This approach has been valid from the Industrial Revolution until the second half of the 20th century. The paradigm of human separation from nature was based on four main assumptions:

1- Humans have cultural and genetic characteristics that distinguish them from other animal species;

2- Cultural and social factors such as technology are major determinants of human differentiation;

3- Cultural and social environments are exhausting substrates for humans, and the biophysical environment is very vast and ambiguous;

4- The concept of culture is cumulative, meaning that technical and social advances can be infinite and solve all social problems [65].

Some sociologists sought to make a paradigm shift in sociological studies and criticized the dominance approach, which considered man as an exception to nature. Thus, some sociologists have redefined the new ecosystem paradigm with respect to the impact of the environment on social behavior and the impact of social processes on the environment. They concluded that humans, like other species, interact with nature and are dependent on the global ecosystem. Thus, within social studies, a new perspective emerged that incorporated environmental variables into social research and provided a new sociology perspective. Considering human initiative, this approach believed that humans are interdependent with other species in the ecosystem. This approach also considered social and cultural forces as an important factor. It should be noted that humans are also affected by ecosystem feedback. Sociologists believed that human society was an integral part of life, but human communities' domination over nature leads to self-destruction [65].

The construction of complex societies by humans has led to the overuse of dependent resources. In fact, human success has been accompanied by the sad story of the destruction of nature. The new ecosystem approach changed previous divisions in sociological theories. This new approach advocated a view that was less human-centered and more ecological. This means that humans are just one of the species living on Earth [65]. This paradigm is based on four main assumptions:

- Humans have exceptional characteristics. They are one species among other species and are interdependent with the global ecosystem.

- Human life is not only influenced by cultural and social factors, but also affected by complex factors such as the causes, effects, and feedback of the life cycle. Therefore, human activities may have unintended consequences.

- Humans are dependent on the biophysical environment. It imposes biological and physical constraints on them.

- Although with human initiatives or factors related to those, it possible to overcome restrictions, but they should in no way violate ecological laws [65].

They took an approach (new ecological paradigm) and insisted that the growing environmental crisis has consequences not only for nature but also for human society. This approach introduces a new perspective in sociology that considers the environment for understanding social conditions as necessary, much as political and economic processes.

Environmental value includes a person's basic opinions about the environment and reflects their worldview. Studies indicate that people who have more biological values (have intrinsic respect for the environment) usually pay more attention to the environment than those with technological values (believe that technology can provide solutions to environmental problems) [67].

Knowledge is considered a necessity for achieving success. It is used as a tool to overcome psychological barriers (such as ignorance or misinformation). Although knowledge does not always directly affect behavior, it reinforces other features that facilitate behavior change [68].

Some researchers care a lot about environmental knowledge and believe that there is a broad and universal knowledge on biological and ecological concepts to restore and maintain environmental quality [15]. Another group of researchers defines environmental knowledge as an individual's ability to understand and evaluate society's impact on ecosystems and point out that environmental knowledge means understanding environmental issues, their problems, and their consequences. However, some other sociologists in their studies have proposed three forms of environmental knowledge, which are: systematic knowledge, action knowledge, and effective knowledge. Systematic knowledge deals with how ecosystems work. An example of systematic knowledge is the relationship between carbon dioxide and global climate change. Knowledge of action encompasses a wide range of behavioral strategies (such as activities related to reducing carbon dioxide). Effective knowledge is the knowledge that helps a person to choose behavioral solutions. It can be said that real knowledge about the environment is a prerequisite for an environmental attitude [69].

## 3. Customers' sensitivity to electricity prices and their consumption

Customers' reactions to the implementation of various DR programs are different. Therefore, many mathematical models have been proposed to estimate customer reactions. DR models can be presented using the concept of demand-price elasticity [70]. Demand-price elasticity is defined as the sensitivity of the customers to change their load levels in response to the changes of the electricity prices [71], [72]. In some cases, changes in demands relative to prices have been be considered as linear functions [70]; but sometimes its changes can be considered non-linear [73]. In addition to the previous cases, the attitudes and behavior of customers, as well as the implementation of the DR program at different times, can be considered. By proposing a flexible model, the role of time can be included in the model [74]. Some of the DR models are based on optimization methods [75]. The primary assumption in this type of models is customers' logical behavior, and the objective function of the problem could be defined as the maximization of the desirability or minimization of the costs.

In psychology and behavioral economics, non-financial interventions can be as effective in changing consumer consumption patterns as financial incentives [76]. In this regard, special attention can be paid to the energy policy. Rewards and penalties are two strategies that can change the consumption pattern of customers [71]. Although both of these strategies can lead to behavior change in the short term, research has shown that rewards can also lead to habit formation [77]. Based on this theorem, even if long-term incentive DR programs are no longer implemented, it is predicted that most of the positive effects of those programs will remain due to habit formation in customers [78]. Many researchers have also studied the effect of electricity prices on customer behavior to reduce energy consumption. The results show a close relationship between financial incentives and customers' behavioral characteristics [79],[80].

Based on the studies conducted in human behavior, it can be concluded that the "customers' responses to penalties and rewards are the same" assumption is not correct [81]. In fact, this is contrary to the assumption that customer behavior is linear in classical economics. In order to deal with this problem, it is possible to model different effects of customers' reactions to the implementation of two different types of DR programs and use the concept of loss aversion, which is one of the well-known models of human behavior in psychology [82]. The concept of loss aversion implies that people generally prefer to lose less rather than gain [83]. In other words, in financial-behavioral discussions, the fear of loss is a much stronger stimulus than the temptation to profit. In most psychological studies, the fear of loss is almost twice as strong as the temptation to profit [84].

According to behavioral economics research, the effect of loss on a person's inner sense is much stronger than profit [85]. Therefore, when the desirability function is plotted, the curve slope is greater in the loss area than in the profit area. In theories of micro and classical economics, as well as rationally in individuals, it is always assumed that an equal amount of profit or loss causes the same amount of motivation. But considering the characteristics of human behavior, such an assumption is wrong. Figure 1 illustrates an outline of the desirability function [82]. The amount of loss aversion is different in high-risk and low-risk conditions. It should be noted that the slope of the desirability function curve, from which the concept of loss aversion is derived, is higher in high-risk conditions.

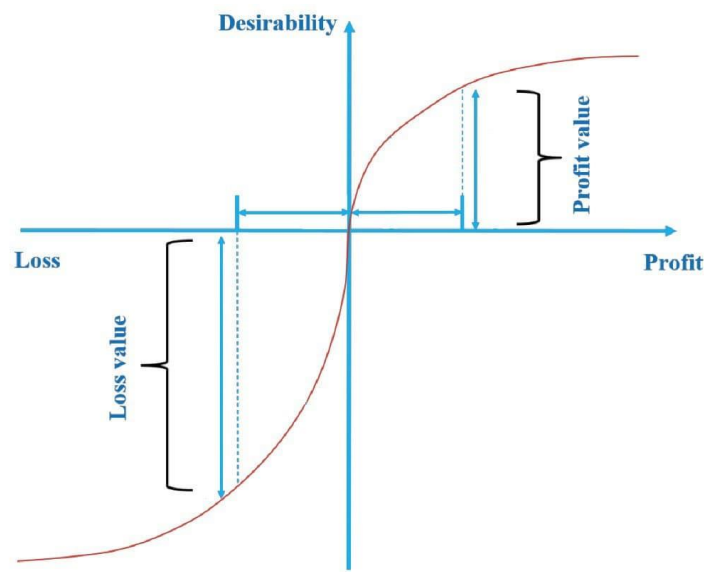

**Figure 1.** Outline of the desirability function [82]

## 4. Analytical model

Based on the issues raised in the previous section, which is derived from the research's theoretical framework, the analytical model of the research is expressed in this section. Based on this information, it can be checked that:

- How cultural, economic, and social capital, as concepts derived from Bourdieu's theory[1] of consumption and lifestyle, influence electricity consumption behavior?
- How people's knowledge of electricity affects electricity consumption behavior?
- How environmental value influences electricity consumption behavior by relying on Kaiser Permanente?
- How the environmental attitude derived from Dunlop's New Environmental Paradigm (NEP) influences electricity consumption behavior?

Based on the studies, there is a significant relationship between the studied variables (such as cultural capital, economic capital, social capital, environmental attitude, environmental value, attitudes toward subsidies, age, employment status, marital status, news resources, social trust, social participation, environmental knowledge, and building interior architecture) and electricity consumption behavior and can be generalized to the statistical population. Also, the behavior of electricity consumption in terms of (employment status, marital status) has a statistically significant difference and can be generalized to the statistical community. Accordingly, people with full-time employment and married people have better electricity consumption behavior than other people [35], [51], [86].

It is expected that the values and attitudes of people towards environmental issues will affect their electricity consumption behavior. They are positive values and attitudes towards environmental issues that will lead to positive changes in the energy consumption patterns. If people of a society value a subject and realize its importance (like the matter of energy and the proper consumption pattern), they will be more cautious about it. Also, positive values about the environment will create a positive attitude towards energy and its adequate consumption [84].

Regarding the effect of the age variable on electricity consumption behavior, it should be said that age can play a decisive role in behavior, lifestyle, and consumption pattern for several reasons. People of different ages have different needs and have different abilities to meet their needs. Also, their experiences of the past years would affect the lifestyles and consumption patterns.

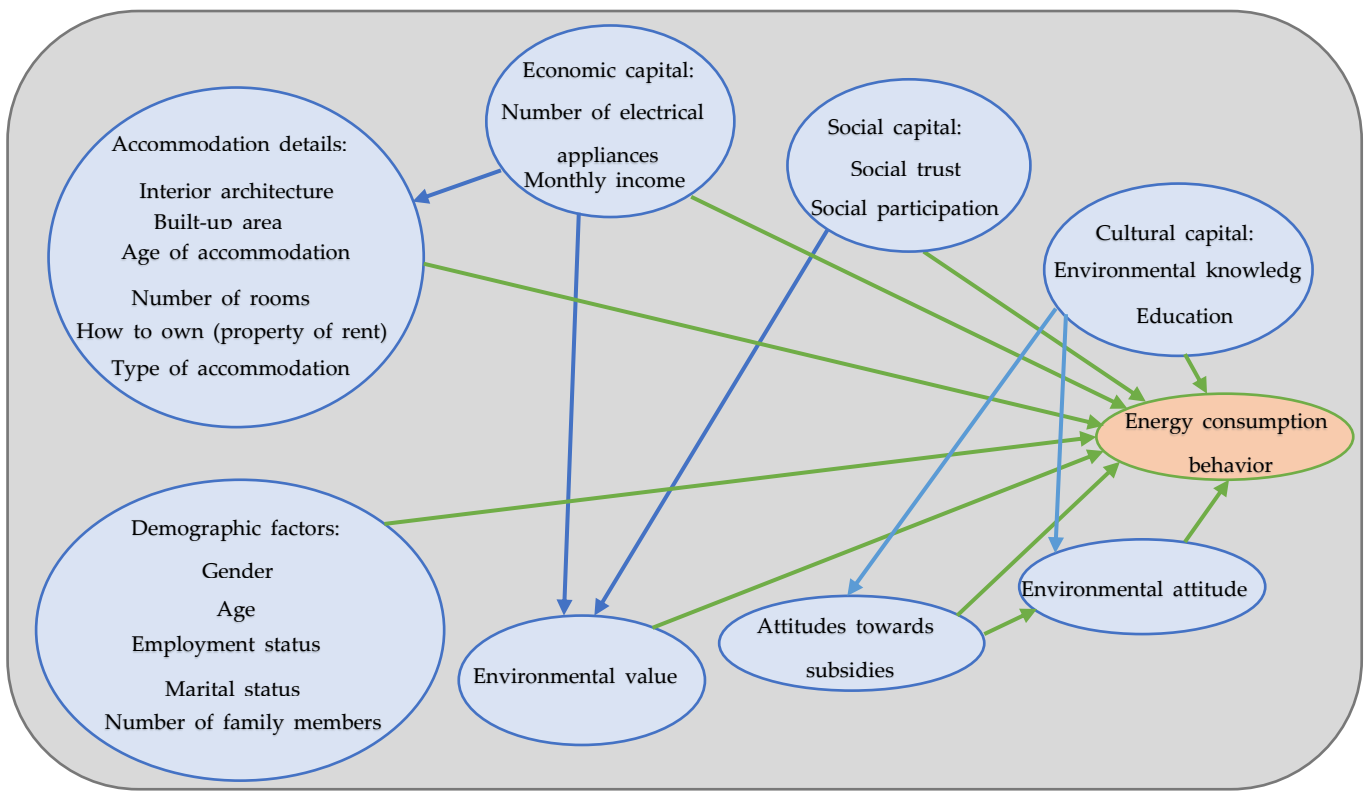

**Figure 2.** Analytical model [65]

---

[1] Bourdieu's theory defines lifestyle as a set of systematic activities that arise from the verve of each individual and have more of an objective and external aspect. According to Bourdieu, lifestyle identifies the individual personality symbolically and also distinguishes between different social strata [87].

One of the critical variables affecting energy consumption behavior is the knowledge about environmental issues, which was examined as one of the dimensions of cultural capital. Knowledge is a factor that affect people's environmental behavior. Consumer behavior should be promoted towards conscious and responsible manner by increasing the citizens' awareness of environmental issues related to energy consumption. Therefore, having more cultural capital means having a higher cognitive ability. In addition, education is another part of cultural capital. As education increases, people become more responsible about consuming electricity. In the process of education, knowledge increases by increasing awareness, and as a result, people will behave more responsibly [9].

Home architecture is also one of the important factors affecting energy consumption behavior. One of these solutions to preserve the limited natural resources is the sustainable design of green buildings, which should prevent energy losses in and provide energy reuse. Attitude toward subsidy schemes is another variable that can be included in the final model. This variable indicates that the more positive the attitude of people towards subsidies, the more appropriate and efficient behavior in electricity consumption will be.

Therefore, a model can be proposed that can show the relationship between these factors. This model is shown in Figure 2.

In this paper, a number of hypotheses have been considered based on the study of resources and interviews with experts. Figure 3 shows the relationship between electricity consumption and various cultural, social, and economic factors. Ten of these factors are directly related to electricity consumption. The further moving to the right of this figure, the more significant these factors' impact on improving electricity consumption compared to the previous factors. For example, the more people follow news related to electricity (such as electricity prices) from more reputable news sources, the more appropriate that person's electricity consumption behavior will be. As another instance, as much as people's employment status will be better or go toward full-time employment, their electricity consumption behavior at home becomes so much better (because they are fewer hours at home). Nevertheless, for example, if we want to compare the influence of these three factors, we see that the effect of the age factor is greater than the factor of news resources. Although, the impact of both these factors is much less than the impact of employment status on improving electricity consumption behavior.

The same is true for factors that are inversely related to electricity consumption. This means that as the amount of these factors increases, electricity consumption status becomes more inappropriate. So, as we move to the left, the factors' impact on getting electricity consumption worse becomes more potent than the previous elements.

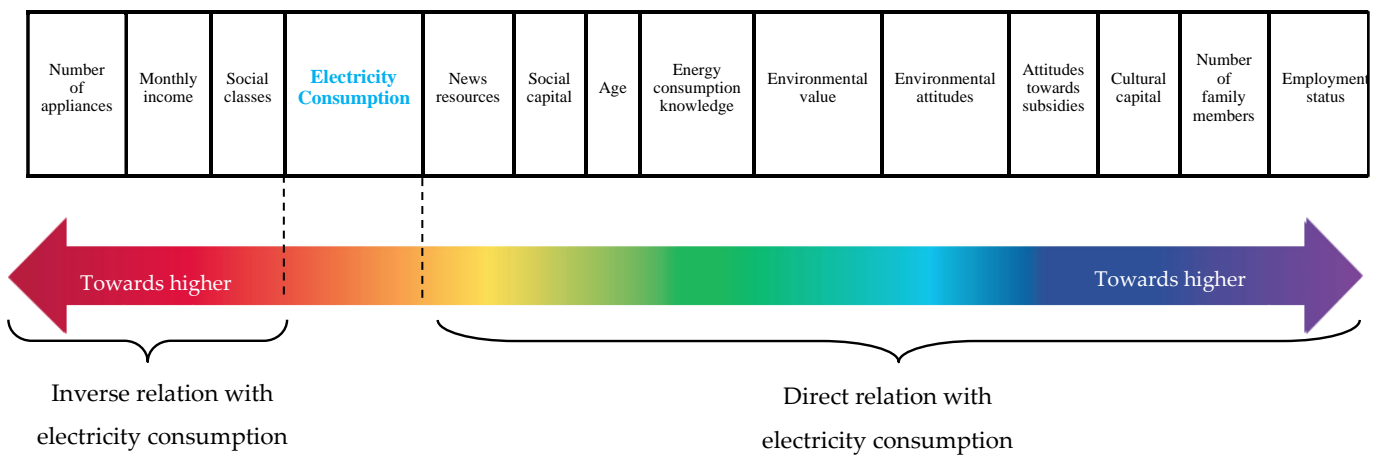

**Figure 3.** Relation between Electricity consumption and different factors

## 5. Discussion, suggestions, and solutions

*5.1. Discussion*

Nowadays, scientists widely discuss global issues such as political and economic crises, fossil fuel sustainability, environmental concerns, increasing population congestion, and economic growth to find appropriate solutions for energy problems in ecological crises. Therefore, one of the most important environmental issues closely related to human life is the energy used in various forms, including water, electricity, and gas.

Excessive reliance on fossil fuel-based resources, excessive energy consumption, and environmental pollution can be among the most critical issues related to people's environmental behaviors [67].

To make changes in consumption, especially in fossil resources, which is always one of the main problems, it is necessary to identify and study the factors affecting energy consumption. Considering the importance of energy consumption and humans' role in its process, this paper examined the factors affecting electricity consumption behavior and some of the requirements for implementing DR programs. For this purpose, first, an analytical model was introduced to present these requirements. Then, the hypotheses about the cultural and social needs for implementing DR programs have been proposed.

*5.2. Suggestions and solutions*

Consumption plays an important role in determining the type, amount, and form of production and distribution. The conditions and culture of society also influence it. If society's situation is such that it leads people to consume more and more, most of the resources of society will be allocated to consumption. In this case, the savings level will be reduced, and there will be no suitable ground for investment. Energy management is one example of consumption pattern modification, and one of the consumption pattern modification ways is DR programs that require the participation of different types of customers. To change people's consumption patterns and optimize them, an appropriate DR program could be selected according to people's behaviors to make it more effective. Besides, people's trusted items should also be taken into account.

Several factors influence the issue of optimizing energy consumption in the household sector in society. Meanwhile, cultural and social characteristics and their impact on consumption patterns and consumption behavior of urban households are becoming more important in urban communities; because social and cultural activities in the urban society are more widespread than in rural society.

According to society's characteristics, responsible behaviors to protect natural resources and the environment should be promoted among society members. This is not achieved without changing the knowledge, value, and attitude of humans towards the environment. Education can indirectly but more effectively influence people's behavior to create an appropriate energy consumption pattern by emphasizing environmental values and enhancing friendship with the environment. The attitude and behavior of consumers can be changed by informing people about some important issues such as the limitations of energy resources, the state of energy consumption in the world, providing solutions to reduce energy consumption, energy management, information about the crisis and possible shortages; Because many undesirable behaviors are due to lack of knowledge of the outcome of the desired behavior and how to achieve it. Therefore, providing information in various ways, including teaching directly, such as brochures, CDs, and face-to-face, as well as indirect teaching, will be very effective in managing energy consumption.

Education is the key issue and with the least cost in this sector, we can achieve great changes in terms of environmental behaviors in society. The media can play a prominent role in motivating people to cultivate a spirit of solidarity, harmony, participation, and a sense of social responsibility. Therefore, social participation growth can positively and effectively impact the pattern of citizens' energy consumption. The media can help get people involved in government policies. They can also make people more aware; and help implement policies by describing the process of energy supply, the amount of energy available, and the sustainable use of energy.

Finally, considering the important factors in people's social trust and their source of information, education can be mentioned as one of the main requirements for DR, which can be expressed in different formats and according to the proposed factors.

**6. Concluding remarks and outlook**

This paper examines the factors affecting the behavior of electricity consumption and the requirements for the implementation of demand response programs from a cultural, social, and behavioral perspective. For this purpose, first, according to the theories among sociologists, an analytical model was introduced to introduce these needs. Then, the hypotheses about the cultural and social needs for the implementation of demand response programs are proposed. The issue of optimizing energy consumption in the household sector is influenced by several factors and components. Meanwhile, the position of cultural and social components and their impact on the consumption pattern and consumption behavior of urban households are becoming more important and visible. Variables that contribute to the explanation of electricity consumption behavior are social participation, environmental value, environmental knowledge, consumer age, attitudes toward subsidies, home interior design, and environmental attitudes. Considering the important factors in the field of social trust of the people and their source of information, the provision of education was mentioned as one of the main requirements for demand response program participation, which can be expressed in different media.

Finally, it is concluded that in order to protect natural resources and the environment, responsible behaviors should be promoted among the members of society. This cannot be achieved without changing human knowledge, value, and attitude towards the environment. Education can indirectly and more effectively influence people's behavior to create an appropriate energy consumption pattern by emphasizing environmental values and strengthening the sense of environmental friendliness.

Informing people about energy resource constraints, the state of energy consumption in the world, providing information on ways to reduce energy consumption, informing about crises and possible shortages is an important part of changing the attitudes, values, and behavior of consumers. Providing information in various ways, including direct instructions, face-to-face classroom instructions, and indirect instructions (such as CDs, brochures, pamphlets and posters) will effectively manage energy consumption.

Education is a key factor that could provide significant changes in environmental behaviors, with low expenditure requirements. Media can also engage people with governmental policies by informing them about the goals and plans, and explaining the energy supply process and the amount of energy available. In addition, the building sector is one of the largest sources of energy consumption in most communities. Therefore, paying attention to building technologies to save and optimize energy consumption can play an influential role.

More complementary researches could be done in the future to identify the effective variables (such as of economic capital and social classes). Since effective variables could completely be depended to the specifications of the system under study and citizen's habits and beliefs, several local studies could be conducted to identify and investigate the effective factors to form proper consumption patterns.

**Acknowledgments**

This work is supported by *Niroo Research Institute* (NRI) under Contract No. 179101. Financial supports granted by Niroo Research Institute (NRI) are gratefully acknowledged.